\documentclass[aps, prb,superscriptaddress, twocolumn,showpacs,preprintnumbers,amsmath,amssymb]{revtex4}

\usepackage{array}
\usepackage{amsmath}
\usepackage{amssymb}
\usepackage{epsfig}
\usepackage{graphicx}
\usepackage{wasysym}
\usepackage{bbm}

\begin{document}
\title{Weak indices and dislocations in general topological band structures}

\author{Ying Ran}
\affiliation{Department of Physics,
Boston College, Chestnut Hill, MA 02467, USA}
\date{\today}

\begin{abstract}
It has recently been shown that crystalline defects -
dislocation lines - in three dimensional topological insulators, can host
protected one dimensional modes propagating along their length. We
generalize this observation to the case of topological superconductors and
other insulators of the Altland Zirnbauer classification, in d=2,3
dimensions.  In general, protected dislocation modes are controlled by the
topological indices in (d-1) dimensions. This is shown by relating this
feature to characteristic properties of surface states of these
topological phases. This observation also allows us to constrain these
surface states properties, which is illustrated by an addition formula for
(d-1) and (d-2) indices of a topological superconductor.
\end{abstract}

\maketitle

Topological
insulators, a new phase of matter with protected surface
states have attacted strong interest both theoretically and experimentally (for reviews, see \cite{hasan_kane_review,Moore_Review_2010,qi:33}). Recently we showed that dislocations in a three-dimensional topological insulator can host one-dimensional helical modes traveling along its length\cite{disloc_nature}. In addition, 3D topological insulators are characterized by four $Z_2$ topological indices: one ``strong index'' and three ``weak indices''\cite{roy-2006,fu:106803,moore:121306}. The weak indices are related to the strong index in one lower dimension: two spatial dimension, and we will also call them (d-1) indices. We showed that it is these (d-1) indices that controls the existence of one-dimensional helical modes hosted by dislocations\cite{disloc_nature}. 

3D topological insulator is just one example of many different types of topological band structures\cite{PhysRevB.78.195125,Kitaev_Periodic_Table}. Can these results be generalized to other topological band structures? Do crystalline defects in other topological band structures host topological bound states? Are there (d-1) indices in general topological band structures? Is the existence of these bound states also governed by (d-1) indices? Our answers to these questions are positive. 

In this paper we discuss dislocations and lower-dimensional indices in general topological band structures. If we require the topological insulator or superconductor to have stable edge states (stable in the presence of impurities), there are generally 10 classes of them\cite{PhysRevB.55.1142,Mehta_Random_Matrices} according to the
presence or absence of the time-reversal, particle-hole and chiral symmetries. One can classify these 10 classes of topological band structures in the presence of disorder (and in the absence of crystalline order) by, for example, studying the localization problem on the surfaces\cite{PhysRevB.78.195125,Kitaev_Periodic_Table}. The resulting classifications are labeled by the so-called strong topological indices in each dimension. We copy the table of general topological band structures from literature\cite{PhysRevB.78.195125} in the following:

\begin{widetext}
\begin{tabular}{|c|c||c|c|c||c|c|c|}\hline
 &  & TRS & PHS & SLS & $d=1$ & $d=2$ & $d=3$ \\ \hline
standard & A(unitary) & 0 & 0 & 0 & - & $Z$ (IQH) & - \\ \cline{2-8}
(Wigner-Dyson) & AI(orthogonal) & +1 & 0 & 0 & - & - & - \\ \cline{2-8}
× & AII(symplectic) & $-1$& 0 & 0 & - & $Z_2$  & $Z_2$ \\
× &  & &  &  &  &  (QSH) &  (strong top. ins.)\\
 \hline\hline
chiral & AIII(chiral unitary) & 0 & 0 & 1 & $Z$ & - & $Z$ \\ \cline{2-8}
 & BDI(chiral orthogonal) & $+1$ & $+1$ & 1 & $Z$ & - & - \\ \cline{2-8}
× & CII(chiral symplectic) & $-1$ & $-1$ & 1 & $Z$ & - & $Z_2$ \\ \hline\hline
BdG & D & 0 & $+1$ & 0 & $Z_2$  & $Z$  & - \\
× &  & &  &  & (spinless p-wave) & (spinless $p+ip$)  &  \\ \cline{2-8}
× & C & 0 & $-1$ & 0 & - & $Z$ (Spin QH) & - \\ \cline{2-8}
× & DIII & $-1$ & $+1$ & 1 & $Z_2$ & $Z_2$  & $Z$  \\
× &  & &  &  & (p-wave with TR) & ($p+ip\uparrow$\&$p-ip\downarrow$)  & (He$_3$-B) \\ \cline{2-8}
× & CI & $+1$ & $-1$ & 1 & - & - & $Z$ \\\hline
\end{tabular}
{Table: Complete list of strong indices of topological insulators(superconductors) and their physical realizations.}
\end{widetext}

What are the physical meaning of these strong indices? First they are topological invariants of the bulk band structures. They also characterize the surface(edge) states. For example, the edge states of the quantum hall insulator (class A in two dimension) is characterized by an integer, which is the number of one dimensional chiral modes traveling along the edge. And the surface states of a 3D topological insulator (class AII in three dimension) is characterized by a $Z_2$ number, which is the parity of the number of the surface Dirac nodes. 

On the other hand, as we will show shortly, this characterization is not complete in the presence of crystalline order. For example, the Dirac nodes of a 3D topological insulator (class AII) can be at $(0,0)$ of the surface Brillouin Zone, and it can also be at $(0,\pi)$. The difference of these two possible surface states are not characterized by the strong index. Instead it is the (d-1) indices (or the weak indices) that detemine the location of the Dirac nodes on the surfaces. In Ref\cite{disloc_nature}, we showed that these weak indices also control the existence of the dislocation-hosted helical modes. We will generalize these results from class AII to all classes and dimensions in the above Table, and discuss the new physics of general topological band structures in the presence of the crystalline order.

\begin{figure}
 \includegraphics[width=0.26\textwidth]{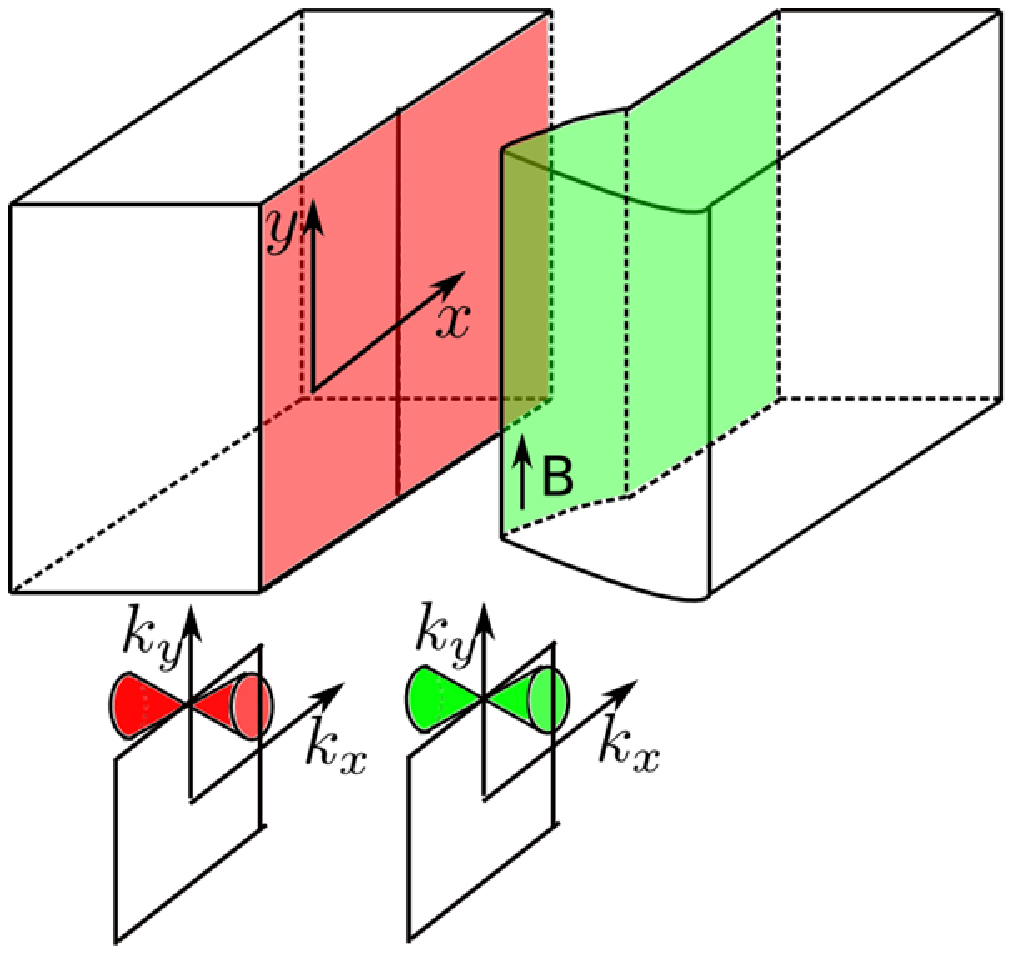}\;\;\;\includegraphics[width=0.2\textwidth]{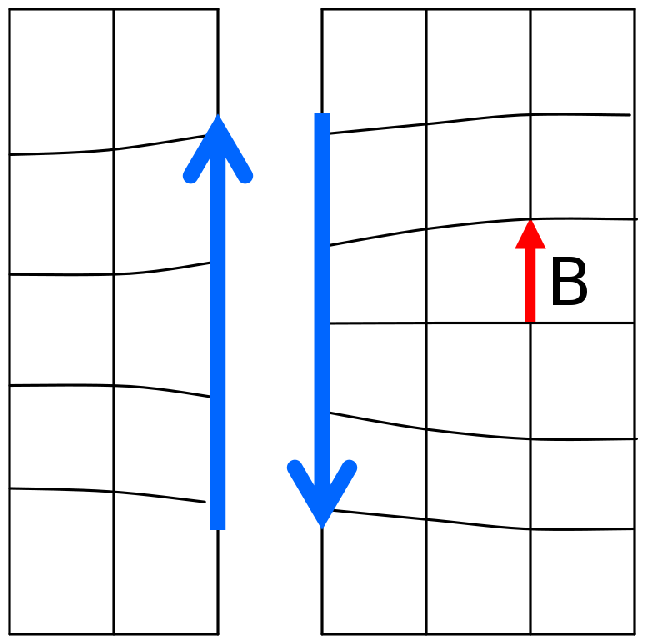}
\caption{(color online) Cut-glue technique in 3D(left) and 2D(right). The surface(edge) states are also shown schematically. $\vec B$ is the Burger's vector of the dislocation.}\label{fig:cut-glue}
\end{figure}

Let us firstly provide some physical intuitions. There is a simple way to understand the existence of the dislocation-hosted helical modes in a 3D topological insulator, which is the cut-glue technique. As shown in Ref \cite{disloc_nature} (see Fig.\ref{fig:cut-glue}), one can create a dislocation by cutting a 3D sample into two halves ($z<0$ and $z>0$ respectively) and then gluing them back nontrivially. After cutting, two surfaces parallel to the $x$-$y$ plane are created. When gluing back, only the $x>0$ part of the two surfaces  are glued back as before, and the $x<0$ part of the two surfaces are glued back after translating by one lattice spacing along $y$ direction. This precedure creates a dislocation along $x=z=0$ line with Burger's vector along $y$ direction. 

After cutting, the surface modes on the two surfaces can be described by Dirac equation. After gluing back, a mass term will be introduced to gap out the Dirac surface modes. In particular, it is clear that the gluing precedure actually introduces a $\pi$ Berry phase for the Dirac cone if it is at $k_y=\pi$. This means the mass term for this Dirac cone changes sign from $x>0$ to $x<0$, namely the dislocation is nothing but a mass domain wall of the surface Dirac equation. It is well-known that such mass domain wall hosts topological pretected gapless modes\cite{CallanHarvey}. Application of such domain-wall-fermion bound states in condensed matter physics has been discussed in various systems such as polyacetylene\cite{PhysRevLett.42.1698,RevModPhys.60.781} and PbTe-type semiconductors\cite{PhysRevLett.57.2967,Boyanovsky1987340}.

A careful study of these domain wall fermions in 3D topological insulators proves that $\vec B\cdot \vec M=\pi $(mod $2\pi$) is the condition to have helical mode inside the dislocation\cite{disloc_nature}, where $\vec B$ is the Burger's vector (real space vector) of the dislocation and $\vec M$ is a momentum space vector composed of the (d-1) indices (also called weak indices in literatures).

This cut-glue technique can easily be generalized to other topological band structures. We simply need to classify the surface states of a general topological band structure in the presence of crystalline order. We find that apart from the possible (d-1) indices, there are also possible (d-2) topological indices. In fact we will show that all the (d-1) and (d-2) indices can be naturally found by classifying the surface states. And then we show that these surface states, after the cut-glue precedure, can host gapless topological bound states via, for example, domain wall fermion mechanism. 

Our main results is the following theorem: \emph{ In the presence of lattice order, the (d-1) indices of a $d$-dimension ($d>1$, otherwise if $d=1$ there is no (d-1) indices.) topological insulator(superconductor) of a given symmetry class,
is composed of the $d$ strong indices, if any, of the $(d-1)$-dimension topological insulator(superconductor) in the same symmetry class. These (d-1) indices form a momentum space vector $\vec M$ and control the existence (or the number) of the stable gapless modes hosted by a dislocation with Burger's vector $\vec B$ via the dot product $\vec B\cdot \vec M$. At the same time, if $d=3$, there are also three (d-2) indices composed of, if any, the $d=1$ strong index in the same symmetry class. These (d-2) indices form a real space vector, and cannot be detected in a generic dislocation defect, but can be detect by the surface modes.} We note that the (d-1) and (d-2) indices are presented as a mathematical results in a remarkable work of Kitaev\cite{Kitaev_Periodic_Table}. \footnote{We only consider the $d=3,2,1$ topological indices. A careful reader may notice that there are also $d=0$ indices in Kitaev's work. Those indices are simply related the total number of filled bands and are high energy properties of the system. For example, by adding a core-level, the $d=0$ index of a band structure can be modified. These $d=0$ cannot modify low energy properties of the system and cannot be detect by the surface modes. We therefore do not discuss these $d=0$ indices.} In this work, we show these indices in a different and more physical fashion. The physical meaning of these indices, in particular the relation of these indices and the surface modes are explicitly shown. The well-known time-reversal symmetric topological insulator (class AII) is just a special case of the above theorem.

We prove this theorem by discussing the ten classes case by case. In the following we only show two cases as examples and the other cases are studied in Appendix\ref{app:general_TI}. These two examples are class D in two dimension and class DIII in three dimension.

Another important result that we obtain is a general addition formula involving both the (d-1) and (d-2) indices: Eq.(\ref{eqn:addition_rule1}). This formula is useful when studying the interface-hosted bound states of two different topological band structures.

\begin{figure}
\includegraphics[width=0.48\textwidth]{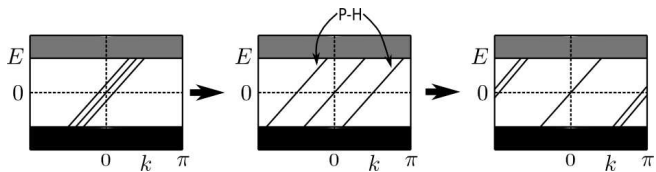}
\caption{Deformations of edge modes of (class D $d=2$) p+ip topological superconductor in the edge Brillouin zones.  The particle-hole symmetry (P-H in figure) allow the motion of a pair of majorana edge modes from $k=0$ to $k=\pi$. The parity of the number of Dirac nodes at $k=\pi$ is a topological invariant.}
\label{fig:deform}
\end{figure}

\textbf{(class D $d=2$)} This is the class of p+ip superconductor. The strong index $\nu_0\in Z$ indicates the number of chiral majorana fermion modes at the edge. Are there other topological invariants of the edge states? Consider the edge of lattice vector $\vec a_1$. By studying the arbitrary deformation of the edge modes, one finds that, by particle-hole symmetry, the parity of majorana fermion modes at $k_1=\pi$ is also a topological invariant (see Fig.\ref{fig:deform}). Let us denote it as $m_1$ (takes value of $0$ or $1$) and study the cut-glue predure as shown in Fig.\ref{fig:cut-glue}. 

After cutting, let us assume that there is a single chiral majorana mode at $k_1=\pi$ on each edge. At low energy these edge modes are described by effective theory at $m=0$:
\begin{align}
 H=\int dx \;\gamma^T(-i\partial_x)\tau_z\gamma+m(x) \gamma^T\tau_y \gamma\label{eq:majorana_Dirac}
\end{align}
where $\tau$'s are Pauli matrices. $\gamma(x)=(\gamma_L(x),\gamma_R(x))$ is the two-component majorana fermion field satisfying $\gamma(x)^*=\gamma(x)$, and $\gamma_L(x),\gamma_R(x)$ are the two counter-propagating edge modes living on the two edges. When we glue the two edges back, the two edges are tunneling to each other and a mass gap $m$ is opened up. Note that this is the only non-vanishing mass term in this majorana theory.

Because the majorana edge modes are both at $k_1=\pi$, the dislocation gives a $\pi$ Berry phase to the Dirac equation and the mass $m$ in Eq.(\ref{eq:majorana_Dirac}) forms a domain wall at the dislocation. Namely $m(x)<0$ when $x<0$ and $m(x)>0$ when $x>0$. Solving Eq.(\ref{eq:majorana_Dirac}) one finds a single majorana mode $\gamma_0\sim \int dx\; e^{-\int_0^x m(x') dx'}(\gamma_L(x)-\gamma_R(x))$ bound to the dislocation. One can easily check $[H,\gamma_0]=0$; i.e., this mode is at zero energy. Note that the majorana edge modes at $k_1=0$ do not see the dislocation as a mass domain wall and thus do not cause majorana bound state.

Because two majorana bound states can mix and open up an energy gap, we conclude that $\vec a_1$-dislocation has a majorana zero mode bound to it if and only if the parity of majorana edge modes at $k_1=\pi$ is odd. Similarly there is another invariant $m_2$ at edge $\vec a_2$ controlling the existence of majorana mode in dislocation with Burger's vector $\vec a_2$. $m_1,m_2$ are the (d-1) indices.

In summary, the (d-1) indices can be written as a momentum space vector $\vec M=(m_1\vec G_1+m_2\vec G_2)/2$, $m_i\in Z_2$. And the condition of the existence of dislocation majorana zero mode is given by $\vec B\cdot\vec M=\pi(\mbox{mod} 2\pi)$. Although we only prove this condition for $\vec B=\vec a_1,\vec a_2$, the result is obviously true for general $\vec B$. This is because a general $\vec B$-dislocation can be thought as a combination of several dislocations with $\vec B=\vec a_1,\vec a_2$.

We only studied the edge states of the $\vec a_1,\vec a_2$ edges. What about the edge states of a general edge along $\vec a_1'=u_1\vec a_1+u_2\vec a_2$ direction? Another way of phrasing the problem is that: is $\vec M$ depends on the choice of the real space basis $\vec a_1,\vec a_2$? In fact we already proved that the answer is negative. This is because $\vec M$ can be detected by dislocations with a general Burger's vector, and cannot depend on choice of basis.

Here we comment on the relation between the current work and previous studies. Kou and Wen studied the translation symmetry protected superconductors\cite{PhysRevB.80.224406}, and they find that there are four $Z_2$ indices corresponding to the parity of the number of electrons on even by even, even by odd, odd by even and odd by odd lattices respectively. These four $Z_2$ indices are related to our results and can be understood as follows: they are composed of the parity of the strong index, the two (d-1) $Z_2$ indices, and one $(d-2)=0$ dimensional index. The last $d=0$ index, as mentioned in the footnote, is nothing but the parity of the total number of filled bands. This index can be modified by simply adding a core level, and cannot be detected by any low energy probe. Therefore we do not study this $d=0$ index in the present study.

\textbf{(class DIII $d=3$)} The strong index $\nu_0\in Z$ indicates the number of majorana Dirac nodes at the surface. Apart from $\nu_0$, there are other topological invariants of the surface states in the presence of crystalline order. Consider the surface spanned by lattice vector $\vec a_1$ and $\vec a_2$. By considering arbitrary deformation of the surface modes, one finds that, by time-reversal symmetry, the parity of the Dirac nodes at $(\pi,0)$, $(0,\pi)$ and $(\pi,\pi)$ are also topological invariants. We thus have three extra invariants at surface spanned by $\vec a_1-\vec a_2$. We re-group the three $Z_2$ invariants as: the parity of the number of Dirac nodes along $k_1=\pi$: $m_1$ ($m_1=0$ or $1$), the parity of the number of Dirac nodes along $k_2=\pi$: $m_2$ ($m_2=0$ or $1$), and a third index $v_{12}$. The definition of $v_{12}$ is trickier: we choose a new surface momemtum space basis $\{\vec G_1',\vec G_2'\}$ such that $G_1'=(m_1 \vec G_1+m_2 \vec G_2)$ (if $m_1=m_2=0$ then $\{\vec G_1',\vec G_2'\}$ can be arbitrary.). We define the parity of the number of Dirac nodes at $\vec G_2'/2$ as $v_{12}$ ($v_{12}=0$ or $1$). One can easily see that this definition of $v_{12}$ is independent of the choice of $\vec G_2'$ as long as $\vec G_1',\vec G_2'$ form a momemtum space basis. We will show that $m_1,m_2$ are (d-1) indices and $v_{12}$ is a (d-2) index. 

By considering the cut-glue procedure in Fig.\ref{fig:cut-glue}, one finds that the parity of Dirac nodes along $k_2=\pi$ controls the existence of majorana helical modes (a pair of counter-propagating majorana modes which respects time-reversal symmmetry.) in a dislocation with Burger's vector $\vec B=\vec a_2$. This can be proved as follows. Assume there is a single majorana Dirac node at $(k_1,k_2)=(0,\pi)$. After cutting, the low energy effective theory of the surface modes is
\begin{align}
 H=&\int dxdy \gamma^T[(-i\partial_x)\tau_x+(-i\partial_y)\tau_z\mu_z]\gamma+m(x)\gamma^T\tau_z\mu_y\gamma,\label{eq:surface}
\end{align}
where $\gamma=(\chi_L,\eta_L,\chi_R,\eta_R)$ are the continuum limit of the majorana fermions living close to $(0,\pi)$ ($L,R$ represent the left and right surfaces). Pauli matrices $\tau$ is in the $\chi,\eta$ space and $\mu$ is in the $L,R$ space. Mass term $m(x)$ represents the tunneling between the two surfaces. Under time reversal transformation, $\chi\rightarrow \eta$, and $\eta\rightarrow -\chi$ as required by Kramer's theorem, namely $\gamma\rightarrow i\tau_y\gamma$. Pauli principle requires that the matrices in the kinetic term is symmetric, and the matrix in the mass term is antisymmetric. Finally $H$ is time reversal symmetric which indicates Eq.(\ref{eq:surface}) is the only correct low energy theory up to convention.

These Dirac majorana fermions see the dislocation as a $\pi$ flux tube after gluing back, because they are at momentum $k_2=\pi$ and the Burger's vector $\vec B=\vec a_2$. This means that the mass term changes sign at the dislocation: $m(x)>0$ when $x>0$ and $m(x)<0$ when $x<0$. Such a mass domain wall will have a single pair of counter propagating one-dimensional majorana fermion modes bound to it. This can be shown readily by solving Eq.(\ref{eq:surface}) as follows.

Because $k_y$ is still good quantum number, we first solve Eq.(\ref{eq:surface}) at $k_y=0$. We find two zero energy bound states: $\gamma(k_y=0,x)\sim e^{-\int_0^x m(x')dx'}\phi_{\pm}$, where $\phi_{\pm}$ are the two eigenstates of $\tau_y\mu_y$ with eigenvalue $+1$. Because $\tau_y\mu_y$ and the $k_y$ matrix $\tau_z\mu_z$ commutes, and can be diagonalized simultaneously into $[1,1,-1,-1]$ and $[1,-1,1,-1]$ respectively, we immediately know that $\phi_{\pm}$ have opposite velocity along $y$-direction, which is the dislocation direction. Because these two modes are related by time-reversal symmetry, we name them as majorana helical modes.

We only considered one Dirac node at $(0,\pi)$. In general, any Dirac node at $k_2=\pi$ will contribute a pair of majorana helical modes. If there are even number of pairs, these helical modes will open up a gap by mixing. Thus we conclude that $m_2$ is odd if and only if there is a single pair of majorana helical modes traveling along dislocation with $\vec B=\vec a_2$. Similar statement is true for $m_1$.

We just showed that the two invariants $m_1,m_2$ controls the existence of the dislocation-hosted majorana helical modes with Burger's vector $a_1,a_2$ respectively. Similarly there are four extra invariants at surface spanned by $\vec a_2-\vec a_3$ and $\vec a_3-\vec a_1$. Let us denote them by $m_2',m_3$ (from $\vec a_2-\vec a_3$ surface) and $m_3',m_1'$ (from $\vec a_3-\vec a_1$ surface). But these six invariants are not completely independent. We must have $m_2'=m_2$ because they both are controlling the existence of majarana helical modes bound to an $\vec a_2$-dislocation. Similarly $m_1=m_1'$ and $m_3=m_3'$. We thus have only three independent invariants $m_1,m_2,m_3$. These are the (d-1) indices.

We can further group them into a momentum space vector $\vec M=(m_1\vec G_1+m_2\vec G_2+m_3\vec G_3)/2$, $m_i\in Z_2$, and the condition of the existence of dislocation majorana helical mode is $\vec B\cdot\vec M=\pi(\mbox{mod} 2\pi)$. Note that we only prove this condition when $\vec B=\vec a_1, \vec a_2,\vec a_3$, but the result is obviously true for general $\vec B$. This is because a general dislocation can be viewed as a combination of dislocations with $\vec B=\vec a_1, \vec a_2,\vec a_3$. This also proves that $\vec M$ is independent of the choice of real space basis $\vec a_1, \vec a_2,\vec a_3$.

Now we are going to establish the fact that $v_{12}$ a (d-2) index that is related to the $d=1$ strong index. Imagine we add an extra band by stacking one-dimensional DIII chains along $\vec a_3$-direction, and over $\vec a_1,\vec a_2$ directions. The edge mode of such one-dimensional DIII chain is a Kramer's pair of majorana modes $\chi,\eta$. These edge modes form a two dimension lattice. Under time reversal transformation: $\chi\rightarrow \eta$ and $\eta\rightarrow -\chi$. We can combine them into a complex fermion $c=\chi+i\eta$. Under time reversal $c\rightarrow ic^{\dagger}$ and $c^{\dagger}\rightarrow -ic$. Now we can write down all possible lattice fermion bilinears allowed by time-reversal. The hopping terms are not allowed as $tc_i^{\dagger}c_j\rightarrow t^{*}c_ic_j^{\dagger}=-(tc_i^{\dagger}c_j)^{\dagger}$. Pairing terms are allowed as $\Delta c_i^{\dagger}c_j^{\dagger}\rightarrow -\Delta^{*}c_ic_j=(\Delta c_i^{\dagger}c_j^{\dagger})^{\dagger}$. Thus these $c$-fermions must have a pure pairing hamiltonian on the $\vec a_1-\vec a_2$ surface. In momentum space the hamiltonian is $H=\sum_k\Delta(\vec k)c_k^{\dagger}c_{-k}^{\dagger}+h.c.$. Pauli statistics requires that $\Delta(\vec k)=-\Delta(-\vec k)$ and thus pairing must vanish and hamiltonian generally forms Dirac nodes at the four time-reversal centers: $(0,0)$, $(\pi,0)$, $(0,\pi)$ and $(\pi,\pi)$. These four Dirac nodes have total chirality zero and thus do not modify the strong index in three dimension. And they also do not modify $m_1$ and $m_2$. They only modify $v_{12}$. Because the chains are along $\vec a_3$ directoin, they also do not modify $m_3$, $v_{23}$ and $v_{31}$.

We just showed that $v_{12}$ can be independently modified by adding an extra quasi-one-dimension band. Similarly $v_{23}$ and $v_{31}$ on the other two surfaces can also be modified independently. Thus there are three $Z_2$ (d-2) indices: $v_{12}$, $v_{23}$ and $v_{31}$. In the following we show these (d-2) indices can be summarized by a real space vector: $\vec V=(v_{12}\cdot \vec a_3+v_{23}\cdot \vec a_1+v_{31}\cdot \vec a_2)/2$. And for a surface whose normal direction (a momentum space vector) given by $\vec G$, the condition to have odd number of Dirac nodes at $\vec G_2'$ point on this surface is $\vec G\cdot \vec V =\pi \mbox{(mod $2\pi$)}$.

\begin{figure}
 \includegraphics[width=0.4\textwidth]{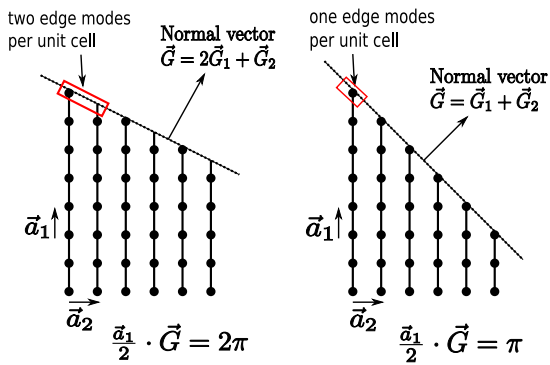}
\caption{Stacked 1D topological superconductor chains along $\vec V=\frac{\vec a_1}{2}$ direction. If the normal vector $\vec G$ of a surface satisfies $\vec V\cdot \vec G=\pi$ (mod $2\pi$) (right) or $\vec V\cdot \vec G=0$ (mod $2\pi$) (left), there will be odd or even number of edge modes per unit cell respectively.}
\label{fig:cut_chain}
\end{figure}

The above statement can again be understood by the stacking limit of 1D DIII chains. We merely need to show: if we have 1D DIII chains along the $\vec V=(v_{12}\cdot \vec a_3+v_{23}\cdot \vec a_1+v_{31}\cdot \vec a_2)/2$ direction stacked and form a 3D lattice, then for an arbitrary surface labeled by normal vector $\vec G$, the chain will terminate on it with a odd number edge modes in a surface unit cell if and only if $\vec G\cdot \vec V =\pi \mbox{(mod $2\pi$)}$. This is obvious, as schematically shown in Fig.\ref{fig:cut_chain}. And if and only if there are odd number of edge modes in a unit cell, when dispersion along the surface are allowed, they form the four Dirac nodes at $(0,0)$, $(\pi,0)$, $(0,\pi)$ and $(\pi,\pi)$. An even number of edge modes in a unit cell will simply open up a full energy gap when tunneling is allowed. These Dirac nodes explains the contribution to, for instance, $v_{12}$ from this quasi-1D band. Therefore for a band structure with both (d-1) index $\vec M$ and (d-2) index $\vec V$, one can view it as a combination of two band structures: a band structure with $\vec M$ only, and a quasi-1D band with (d-2) index $\vec V$ only.

If one combines two general band structures together and ask about the total topological indices, naively one would guess $(\vec M_1, \vec V_1)+(\vec M_2, \vec V_2)=(\vec M_1+\vec M_2, \vec V_1+\vec V_2)$ (the strong indices would obviously add up.). But this turns out to be incorrect. The addition of (d-2) indices actually involves a cross product of the (d-1) indices:
\begin{align}
 &(\vec M_1, \vec V_1)+(\vec M_2, \vec V_2)\notag\\
=&\big(\vec M_1+\vec M_2, \vec V_1+\vec V_2+2\frac{\vec M_1\times\vec M_2}{(2\pi)^2}\big),\label{eqn:addition_rule1}
\end{align}
where we are in the convention that the real space unit cell volume $\vec a_1\cdot \vec a_2\times \vec a_3=1$ and the momentum space unit cell volume is $\vec G_1\cdot \vec G_2\times \vec G_3=(2\pi)^3$. The factor $2$ is because the (d-2) index ((d-1) index) is one half of a real(momentum) space lattice vector.  Eq.(\ref{eqn:addition_rule1}) can be proved in two steps: First physically there must exist an addition rule of indices that makes the $(\vec M, \vec V)$ space an additive group. And one can easily check that the addition rule of Eq.(\ref{eqn:addition_rule1}) indeed makes $(\vec M, \vec V)$ space an additive group. This means that to prove Eq.(\ref{eqn:addition_rule1}), we merely need to show Eq.(\ref{eqn:addition_rule1}) is valid for a set of generators of additive group $(\vec M, \vec V)$, which one can in turn check case by case explicitly by deforming the surface modes.

\begin{figure}
 \includegraphics[width=0.49\textwidth]{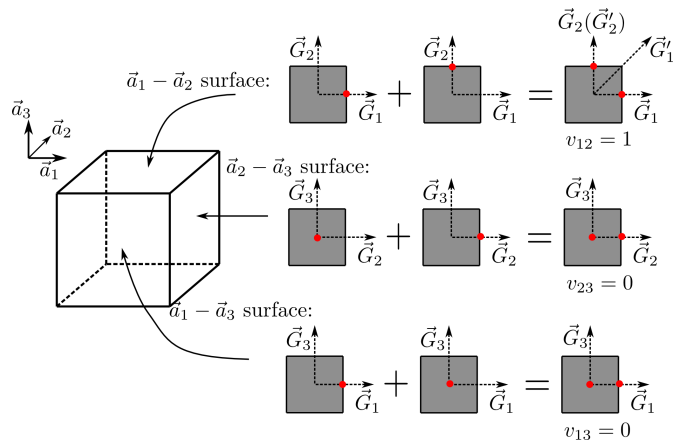}
\caption{(color online) Addition of two band structures with $(\vec M_1, \vec V_1)=(\frac{\vec G_1}{2},0)$ and $(\vec M_2, \vec V_2)=(\frac{\vec G_2}{2},0)$ based on study of surface states. The shaded areas are the surface Brilloin Zones, and the red dots are the surface Dirac nodes. The resulting band structure has $\vec M=\frac{\vec G_1+\vec G_2}{2}$ and $\vec V=\frac{\vec a_3}{2}$.}
\label{fig:addition_rule}
\end{figure}

Here we simply prove Eq.(\ref{eqn:addition_rule1}) for a special case: the two band structures both have $\nu_0=1$ , and $(\vec M_1, \vec V_1)=(\frac{\vec G_1}{2},0)$, $(\vec M_2, \vec V_2)=(\frac{\vec G_2}{2},0)$, and the combined band structure should have $(\vec M,\vec V)=(\frac{\vec G_1+\vec G_2}{2},\frac{\vec a_3}{2})$ as required by Eq.(\ref{eqn:addition_rule1}). This is schematically shown in Fig.\ref{fig:addition_rule}. The readers can easily check the validity of Eq.(\ref{eqn:addition_rule1}) for the more general cases.

Eq.(\ref{eqn:addition_rule1}) is general and true not only for class $DIII$ (see Appendix \ref{app:general_TI}). It can be very useful when studying the topological bound states to the interfaces of two topological band structures, such as a grain boundary of a topological superconductor. If the two topological band structures are labeled by $(\vec M_1, \vec V_1)$ and $(\vec M_2, \vec V_2)$, one can imagine to add an auxilary band structure with indices $(-\vec M_2, -\vec V_2)$ to both sides, which will leave the $(\vec M_2, \vec V_2)$ side topologically identical to vacuum. Such an auxilary band structure will not modify the interface modes topologically. We thus conclude that the interface modes is the same as the surface of a band structure whose indices are $(\vec M_1, \vec V_1)-(\vec M_2, \vec V_2)
=\big(\vec M_1-\vec M_2, \vec V_1-\vec V_2-2\frac{\vec M_1\times\vec M_2}{(2\pi)^2}\big)$.

This concludes our discussion of dislocation, (d-1) and (d-2) indices in a general topological band structure. Upon completion of this manuscript, we note that a study by Teo and Kane\cite{Teo_2010} of general discloations in topological band structures. YR thanks helpful discussion with Ashvin Vishwanath, Hong Yao and Dung-Hai Lee. YR is supported by the startup fund at Boston College.

\begin{appendix}
\section{General (d-1) (and (d-2)) indices and dislocation-hosted bound states}\label{app:general_TI}
\begin{figure}
\includegraphics[width=0.48\textwidth]{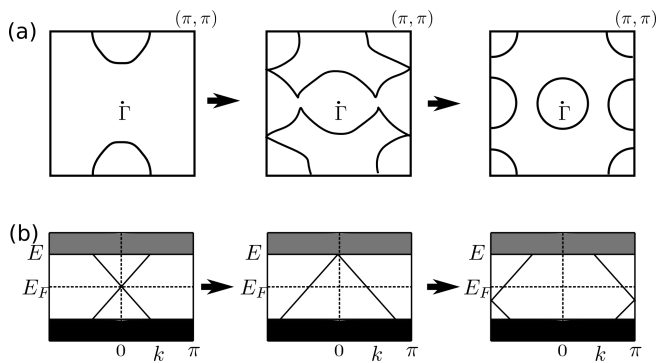}
\caption{Examples of deformations of surface(edge) modes of topological insulators(superconductors) in the surface(edge) Brillouin zones. (a) AII $d=3$: The parities of the numbers of Dirac nodes at $k_x=\pi$ and $k_y=\pi$ are topological invariants. (b) AII $d=2$. A single Dirac node is allowed to move from $k=0$ to $k=\pi$. There is no (d-1) topological invariant.}
\label{fig:deform2}
\end{figure}

\textbf{(class A $d=2$)}: This is the well-known integer quantum hall states. The edge states are characterized by the number of chiral modes, which is an integer (the strong index). As one can easily see, there is no extra topological invariant as one deforms the edge states smoothly. Therefore there is no (d-1) index.

\textbf{(class A $d=3$)}: This is the three dimension integer quantum hall state. We knew that 3D quantum hall states can be understood by a stacking limit of two dimension quantum hall layers and are labeled by a reciprocal lattice vector $\vec G$ that is nothing but the stacking direction of the layers\cite{Halperin_3D_Quantum_Hall}. Here we pretend that we do not know this result and insist on classifying all possible surface states. We will see that we arrive at the same result. However, as we already showed in the main text, our approach is a systematical way of finding topological invariants of band structures and can be easily generalized to other classes where new results can be obtained.

The topologically stable surface states are the chiral fermi surfaces. Given a choice of the surface Brillouin Zone whose basis vectors are $\vec G_1$ and $\vec G_2$, the gapless states on this surface are characterized by two integers $\nu_1$ and $\nu_2$. $\nu_1$ is defined by the number of times that one crosses the chiral fermi surfaces as one moves from $(0,-\pi)$ to $(0,\pi)$ by tuning $k_2$ of the surface momemtum. If the fermi velocity of the chiral fermi surface is positive as it is crossed, it is counted as $1$; and if the fermi velocity is negative, it is counted as $-1$. $\nu_2$ is defined in a similar way by tuning $k_1$ and count the number of times the chiral fermi surfaces are crossed. $\nu_1, \nu_2$ are topological invariants as one cannot change them by deforming the surface states smoothly. And one can easily see that there is no more invariants, namely any two surfaces states with the same $\nu_1, \nu_2$ can be deformed into each other. In addition, we can always make a new choice (denoted as primed) of the surface Brillouin Zone such that $\vec G_1'$ is along $\nu_1\cdot \vec G_1+ \nu_2\cdot \vec G_2$ direction and therefore $\nu'_2=0$ and $\nu_1' G_1'=\nu_1\cdot \vec G_1+ \nu_2\cdot \vec G_2$. These chiral fermi surfaces can then be understood from decoupled layered of atoms stacking in the $G_2'$ direction on the surface, where each layer of atoms has the $\nu_1'$ number of chiral edge modes.

Now if we introduce an edge dislocation whose Burger's vector $\vec B$ parallel with this surface and terminate on this surface. It is clear that this dislocation must host $n=\vec B \cdot \nu_1'  \vec G_1'/(2\pi)$ number of topologically protected chiral modes. This is because the edge dislocation can be simply viewed as the edge of an extra layer of atom with edge modes mentioned above. 

At this moment it seems that the three surfaces each have two integer valued topological invariants. But these six invariants are not independent. One can imagine creating a single dislocation whose Burger's vector parallel with both $xy$-surface and $yz$-surface. The number of chiral modes in this dislocation can be computed from either $xy$-surface or $yz$-surface, and both results must be the same. This gives a constraint between this pair of surfaces and dictates the $\nu_2$ on the two surfaces must be identical. There are totally three such constraints between the three pairs of surfaces. Thus there are three independent topological invariants. These are the three (d-1) indices. And they can be summarized by a three dimension momentum space vector $\vec M=\nu_1\cdot G_1+\nu_2\cdot G_2+\nu_3\cdot G_3$. The number of chiral modes hosted by a dislocation is $n=\vec B\cdot \vec M/(2\pi)$.

\textbf{(class AII $d=2$)} This is class of the Kane-Mele 2D $Z_2$ topological insulator\cite{kane:146802}. The edge modes are classified by a single $Z_2$ index $\nu_0$ which dictates the parity of the number of Dirac nodes. To show there is only one topological invariant $\nu_0$, one can simply make two observations. First $\nu_0$ must be independent of the choice of edge. There are many ways to see this. For example consider a magnetic-superconductivity domain wall\cite{Kitaev_1D_p-wave,Ashvin_Thesis}. It is easy to show that this domain wall hosts a single majorana fermion mode if and only if $\nu_0=1$. But as the domain wall is moved from one edge to another smoothly, the majorana mode cannot disappear. Thus $\nu_0$ is independent of edge. Second one can see that there is no extra topological invariants on a given edge. For example, the time-reversal protected Dirac nodes must be at either $k=0$ or $k=\pi$. One may guess that the parity of the number of Dirac nodes at $k=0$ or $k=\pi$ themselves are topological stable. But one can smoothly move a Dirac node from $k=0$ to $k=\pi$ as shown in Fig.\ref{fig:deform2}.

\textbf{(class AII $d=3$)} This is the time-reversal symmetric topological insulator in three dimension and its (d-1) indices has been proven in literatures\cite{roy-2006,fu:106803,moore:121306}. Here we rederive these results using the approach of classifying the surface modes.

Consider the surface spaned by lattice vector $\vec a_1$ and $\vec a_2$ of the 3D topological insulator. The parity (even or odd) of the total number of Dirac nodes on this surface is determined by the strong index. In the presence of lattice structure, the surface Brillouin zone is well defined. By considering arbitrary deformation of the surface fermi surfaces, one finds that, apart from the parity of the total number of Dirac nodes $\nu_0$, the parity of the Dirac nodes along $k_1=\pi$ and $k_2=\pi$ are also topological invariants (see Fig.\ref{fig:deform}). We thus have two extra invariants at surface spanned by $\vec a_1-\vec a_2$. Similarly there are four extra invariants at surface spanned by $\vec a_2-\vec a_3$ and $\vec a_3-\vec a_1$. But these 6 invariants are not completely independent. 

To see their dependence, we consider their consequence in a dislocation. Consider a skew dislocation of Burger's vector $\vec B=\vec a_1$. Following our discussion in literature\cite{disloc_nature}, this skew dislocation can be obtained by the cut-glue procedure by a cutting $\vec a_1-\vec a_2$ surface containing the dislocation. Atoms on one side of the surface are displaced by $\vec B=\vec a_1$ and reconnected. The Dirac nodes at $k_1=\pi$ will see the dislocation as a mass sign-changing domain wall. This dislocation hosts helical mode if and only if the parity of Dirac nodes at $k_1=\pi$ is odd.

At the same time, we can consider the cut-glue procedure to create same dislocation by cutting a $\vec a_1-\vec a_3$ surface through it. We conclude that the parity of Dirac nodes at $k_1=\pi$ on this $\vec a_1-\vec a_3$ surface must be the same as the one on the $\vec a_1-\vec a_2$ surface, otherwise there will be inconsistency on the existence of helical mode inside the dislocation. Thus there is one constraint between $\vec a_1-\vec a_2$ surface and $\vec a_1-\vec a_3$ surface. Similarly there are another two constraints between the other two pairs of surfaces. Finally we conclude that there are three (d-1) indices which can be summarized as a vector $\vec M=(m_1\vec G_1+m_2\vec G_2+m_3\vec G_3)/2$, $m_i\in Z_2$, and the condition of the existence of dislocation helical mode is $\vec B\cdot\vec M=\pi(\mbox{mod} 2\pi)$.

\textbf{(class D $d=2$)} See the main text.

\textbf{(class D $d=3$)} There is no strong index and similar to class A, the topological protected surface modes are chiral (majorana) fermi surfaces. Given a surface spanned by $\vec a_1,\vec a_2$, we can also classify the chiral majorana fermi surfaces by the two integers $\nu_1,\nu_2$ by the number of times the fermi surfaces are crossed as one tune $k_2$ and $k_1$ from $-\pi$ to $\pi$ respectively. We can also choose a new basis $\{\vec G_1',\vec G_2'\}$ of the Brillouin Zone such that $\nu_2'=0$ and $\nu_1' \vec G_1'=\nu_1\vec G_1+\nu_2\vec G_2$. And again the surface modes can be viewed as the stacking of layers of atoms along the $\vec G_1'$ direction, the edge mode of each layer is $\nu_1'$ chiral majorana modes. But here, in addition to those topological invariants, because we are dealing with a superconductor with particle-hole symmetry, \emph{the parity of the number of chiral majorana fermi surfaces cross $\vec G_2'/2$  is also a topological invariant}. We denote this parity by a $Z_2$ number $v_{12}$. One can easily see that there is no extra topological invariant on this surface. Similarly one can define $\nu$'s on the other surfaces and $v_{23}$ and $v_{12}$.

The $\nu$'s are the (d-1) indices and they control the number of chiral majarana fermion modes hosted by a dislocation. Imagine an edge dislocation terminate on surface $\vec a_1-\vec a_2$ with Burger's vector $\vec B=\vec a_1'$. Because the surface modes can be viewed by the stacking limit, and the edge dislocation can be simply viewed as the edge of an extra disc of atoms, we immediately conclude that this dislocation host $\nu_1'$ number of chiral majorana modes. In general an dislocation whose burgers vector lies within this surface hosts $\vec B\cdot (\nu_1\vec G_1+\nu_2\vec G_2)/(2\pi)$ majorana chiral modes.

Now imagine a dislocation whose Burger's vector $\vec B=\vec a_1$ that connects the surfaces $\vec a_1-\vec a_2$ and $\vec a_1-\vec a_3$. The numbers of chiral majorana modes can also be computed by the two surface must be identical. Thus the $\nu_1$ on $\vec a_1-\vec a_2$ and $\vec a_1-\vec a_3$ must be same. One can therefore conclude that there are totally three (d-1) indices which can be summarized into a single momentum space vector $\vec M=\nu_1\cdot\vec G_1+\nu_2\cdot\vec G_2+\nu_3\cdot\vec G_3$. The number of chiral majorana modes hosted by a dislocation is $n=\vec B\cdot \vec M/(2\pi)$.

The $v$'s are the (d-2) indices. In the following we show that they are related to the $d=1$ strong index and form a real space vector $\vec V=(v_{12}\cdot \vec a_3+v_{23}\cdot \vec a_1+v_{31}\cdot \vec a_2)/2$. One can study an extra quasi-one-dimensional band, whose one-dimensional chain direction is along $\vec V$. The edge mode of a single one-dimensional chain is a majorana mode $\chi$. When these chains are stacked together, the majorana modes can have dispersion on the surface. The most general lattice model of the $\chi$'s, in the momentum space, will be $H=\sum_k \epsilon(\vec k)\chi_k^{\dagger}\chi_{k}$ where $\epsilon(\vec k)$ is real.  Because $\chi_k=\chi_{-k}^{\dagger}$, the Pauli statistics requires that $\epsilon(\vec k)=-\epsilon(-\vec k)$. Thus the surface modes must form fermi surfaces (zero of the real function $\epsilon(\vec k)$) that crosses the inversion centers $(0,0)$, $(\pi,\pi)$, $(0,\pi)$ and $(\pi,0)$. Such surface mode will not have the (d-1) index, but will have non-trivial (d-2) index. One can also easily see that the (d-2) indices of a surface labeled by its normal vector $\vec G$ is non-trivial if and only if there are odd number of majorana edge modes per unit cell, which means nothing but $\vec G\cdot\vec V=\pi (\mbox{mod} 2\pi)$. 

We thus proved that the (d-2) indices can be understood as contributed from a quasi-1D band and the (d-2) indices is nothing but the real space 1D chain direction. Now the 3D class D band structure's (d-1) and (d-2) indices can be grouped as $(\vec M, \vec V)$. And it can be viewed as the combination of two band structures $(\vec M, 0)+(0,\vec V)$. It is important to note that the surface states of $(\vec M, 0)$, whose band structure is the same as the stacking of 2D layers without 1D indices along $\vec M$, gives the \emph{identical (d-1) indices and vanishing (d-2) indices} on any surface as $(\vec M, \vec V)$. Thus the (d-2) indices is completely contributated by the $(0,\vec V)$ part, which is nothing but the stacking of 1D chains. We thus proved the (d-2) index of an arbitrary surface is indeed given by $\vec G\cdot\vec V/\pi$.

We also find that in this case the rule of combining two general band structures is similar to Eq.(\ref{eqn:addition_rule1}):
\begin{align}
 &(\vec M_1, \vec V_1)+(\vec M_2, \vec V_2)\notag\\
=&\big(\vec M_1+\vec M_2, \vec V_1+\vec V_2+\frac{1}{2}\frac{\vec M_1\times\vec M_2}{(2\pi)^2}\big),\label{eqn:addition_rule}
\end{align}
where the only modification from Eq.(\ref{eqn:addition_rule1}) is the factor $1/2$. This is simply because, in our convention, the (d-1) index $\vec M$ is a full reciprocal lattice vector in the present case. Thus adding together two (d-2)-index-absent band structures together may generate a non-trivial (d-2) index. The proof of Eq.(\ref{eqn:addition_rule}) can also be carried out by studying case by case of the a few generators of $(\vec M,\vec V)$ space.

\textbf{(class DIII $d=2$)} The strong index $\nu_0\in Z$ indicates the parity of the number of majorana Dirac nodes at the edge. 
Consider the edge spaned by lattice vector $\vec a_1$. By considering arbitrary deformation of the edge modes, one finds that, by time-reversal symmetry, the parity of the Dirac nodes at $k_1=\pi$ is also a topological invariant. And a dislocation of Burger's vector $\vec B=\vec a_1$ has a Kramer pair of majorana zero modes if and only if this parity is odd. Similarly the parity of Dirac nodes at $k_2=\pi$ for the $\vec a_2$ edge is also a (d-1) index. In summary, the (d-1) indices can be written as a momentum space vector $\vec M=(m_1\vec G_1+m_2\vec G_2)/2$, $m_i\in Z_2$. And the condition of the existence of dislocation majorana Kramer pair is given by $\vec B\cdot\vec M=\pi(\mbox{mod} 2\pi)$.

\textbf{(class DIII $d=3$)} See the main text.

\textbf{(class AI $d=2,3$)} There is no topologically protected edge(surface) modes.

\textbf{(class C $d=2,3$)} This class represents a superconductor with full $SU(2)$ spin rotation symmetry but no time-reversal. When $d=2$ this is the spin quantum hall effect \cite{PhysRevB.60.4245}(basically the spin quantum number plays the role of charge in a usual quantum hall effect). Spin rotation symmetry requires even branches of the chiral mode on the edge. One can easily check that there is no extra topological invariants. So when $d=2$ one only has strong index $n$ where $2n$ is the number of branches of chiral edge modes.

When $d=3$ we should classify all possible chiral fermi surfaces that respect particle-hole symmetry and spin rotation symmetry. Similar to the discussion of 3D integer quantum hall, one can convince oneself that the (d-1) indices is given by three integers and can be summarized into a momentum space vector $\vec M=\nu_1 \vec G_1+\nu_2 \vec G_2+\nu_3 \vec G_3$. Here, for example, $2\nu_1$ is the number of times that the chiral fermi surfaces are crossed as the momentum is tuned from $k_2=-\pi$ to $k_2=\pi$ on the $\vec a_1--\vec a_2$ surface. The band structure can be viewed as the stacking of 2D spin quantum hall layers along $\vec M$ direction. The number of branches of chiral modes hosted by a dislocation with Burger's vector $\vec B$ is $2\cdot\vec B\cdot \vec M$, because the dislocation can be viewed as the edge of a disk of extra atoms, and the factor $2$ is from spin rotation symmetry.

\textbf{(class CI $d=2,3$)} This class is superconductors with full $SU(2)$ spin rotation symmetry and time-reversal symmetry. In $d=2$ one can easily sees that there is no topologically protected edge modes and thus there is no topological band structures.

In $d=3$, as shown in Ref.\cite{PhysRevB.78.195125}, there are topologically protected surface Dirac nodes characterized by an integer $\nu_0$. And the positions of these Dirac nodes are free to move. Thus there is no extra topological invariants.

\textbf{(class AIII, BDI, CII, $d=2,3$)} From now on we discuss the chiral classes AIII, BDI and CII. The common features of these classes is that the strong index in $d=1$ is $\nu_0 \in Z$. This is because there is a chiral symmetry $C$ that is a unitary operator, $C^2=1$,  and $C$ anticommutes with the Hamiltonian. The zero energy states can be labeled by the eigenvalues of $C$, $\pm 1$. If there are two zero modes $c_1,c_2$ with the same eigenvalue, then any bilinear of this two fermions $c_1c_2,c_1^{\dagger}c_2...$ are not allowed by chiral symmetry because they commute with $C$. Imagine we have a 1D chain with $\nu_0=1$ whose left edge is labeled by eigenvalue $+1$ of $C$. Then a ladder of $n$-chains will simply have $n$ copys of zero modes and they do not mix. This ladder has $\nu_0=n$.

The same argument holds if we stack such 1D chains to form 2D or 3D structures. For example, if we stack these 1D chains into a 3D crystal and have a surface cutting the 1D chains. The edge modes of the 1D chains remains at zero energy as one turn on tunneling between the chains because any bilinear is not allowed. Thus for non-interacting system the surface mode contributed by these 1D chains are zero energy flat bands. If we denote the real space full lattice vector along which these 1D chains are pointing along as $\vec V$, the number of zero energy flat bands of a given surface with normal direction $\vec G$ is given by $\vec V\cdot \vec G/(2\pi)$, which is just the number of edges per surface unit cell.

In $d=2$, all three classes have no strong index. The protected edge modes are just these zero energy flat bands. We conclude that the (d-1) indices of these classes are two integers and can be summarized as a momentum space vector $\vec M=\nu_1\vec G_1+\nu_2 \vec G_2$. Here $\nu_1$ and $\nu_2$ are the number of zero energy flat bands along $\vec a_1$ edge and $\vec a_2$ edge respectively. And a dislocation with Burger's vector $\vec B$ host $\vec B \cdot\vec M/(2\pi)$ number of zero modes. This is because an edge along $\vec B$ direction has $\vec B \cdot\vec M/(2\pi)$ number of zero flat bands (for simplicity we assume that components of $\vec B$ does not have common divider, but one immediately sees that the result is general.). Thus the cut-glue precedure immediately gives $\vec B \cdot\vec M/(2\pi)$ zero modes bound to this dislocation. 

In $d=3$, AIII and CII have strong indices characterized by the number of Dirac nodes on the surface. The positions of these Dirac nodes is arbitrary and therefore have no further topological invariants. These three classes have no (d-1) indices. On the other hand, it is possible that the surface hosts zero energy flat bands contributed by the (d-2) index. Based on our discussion, the (d-2) index is a real space lattice vector $\vec V$, and the number of zero energy flat bands of a given surface with normal direction $\vec G$ is given by $\vec V\cdot \vec G/(2\pi)$. 

\end{appendix}

\bibliographystyle{apsrev}
\bibliography{/home/ranying/downloads/reference/simplifiedying}

\end{document}